\documentclass[final]{svjour3} 
\usepackage[english]{babel}
\usepackage[utf8]{inputenc}
\usepackage{amsmath}
\usepackage{amssymb}
\usepackage{mathtools}
\usepackage{tabularx}
\usepackage{subfig}
\usepackage{kpfonts} 
\usepackage{microtype}
\usepackage{booktabs} 
\usepackage{bm} 
\usepackage{listings} 
\usepackage{verbatim} 
\usepackage{color}
\usepackage{hyperref}%
\usepackage{breakurl}
\hypersetup{colorlinks=true,citecolor=blue,linkcolor=blue,urlcolor=blue}

\usepackage{graphicx}

\newcommand{\bigF}{\mbox{\normalfont\Large\bfseries F}}
\newcommand{\bigV}{\mbox{\normalfont\Large\bfseries V}}
\newcommand{\bigzero}{\mbox{\normalfont\Large\bfseries 0}}

\def\makeheadbox{{%
\hbox to0pt{\vbox{\baselineskip=10dd\hrule\hbox
to\hsize{\vrule\kern3pt\vbox{\kern3pt
\hbox{\bfseries [Note:]}
\hbox{This is a peer-review, pre-copyedit version of this article.}
\kern3pt}\hfil\kern3pt\vrule}\hrule}%
\hss}}}

\usepackage{xurl}
\begin{document}

\title{Mathematical Modeling and Investigation on the Role of Demography and Contact Patterns in Social Distancing Measures Effectiveness in COVID-19 Dissemination}

\author{Marco A. Ridenti \and Lara K. Teles \and Alexandre Maranh\~ao \and Vladimir K. Teles }


\institute{M. A. Ridenti
\and Lara K. Teles 
\and A. Maranh\~ao
\at
Aeronautics Institute of Technology, Physics Department, Pra\,ca Marechal Eduardo Gomes, 50 Vila das Ac\'acias, S\~ao Jos\'e dos Campos, SP, Brasil, ZIP code 12228-900 \\
Tel.: +55-12-33058466\\
\email{aridenti@ita.br, lkteles@ita.br} 
\and
V. K. Teles \at
Sao Paulo School of Economics, FGV-SP, Rua Itapeva, 474 Bela Vista, S\~ao Paulo, SP, Brasil, ZIP code 01332-000 \\
\email{vladimir.teles@fgv.br}
}

\date{Received: date / Accepted: date}

\maketitle

\begin{abstract}
\small In this article, we investigate the importance of demographic and contact patterns in determining the spread of COVID-19 and to the effectiveness of social distancing policies. We investigate these questions proposing an augmented epidemiological model with an age-structured model, with the population divided into susceptible (S), exposed (E), infected and asymptomatic (A), hospitalized (H), infected and symptomatic (I), and recovered individuals (R), to simulate COVID-19 dissemination. The simulations were carried out using six combinations of four types of isolation policies (work restrictions, isolation of the elderly, community distancing and school closures) and four representative fictitious countries generated over alternative demographic transition stage patterns (aged developed, developed, developing and least developed countries). We concluded that the basic reproduction number depends on the age profile and the contact patterns. The aged developed country had the lowest basic reproduction number ($R0=1.74$) due to the low contact rate among individuals, followed by the least developed country ($R0=2.00$), the developing country ($R0=2.43$) and the developed country ($R0=2.64$). Because of these differences in the basic reproduction numbers, the same intervention policies had higher efficiencies in the aged and least developed countries. Of all intervention policies, the reduction in work contacts and community distancing were the ones which produced the highest decrease in the $R0$ value, prevalence, maximum hospitalization demand and fatality rate. The isolation of the elderly was more effective in the developed and aged developed countries. The school closure was the less effective intervention policy, though its effects were not negligible in the least developed and developing countries. 

\end{abstract}

\section*{Declarations}

\subsection*{Funding}

Not applicable.

\subsection*{Conflicts of interest}

Not applicable.

\subsection*{Availability of data and material}

The results of the computation and simulation reported here are available in the open GitHub repository \url{https://github.com/mridenti/COVID-DEMOGRAPHY}. We kindly ask that eventual users of the data and material in the repository cite this article and the repository address on any publication that use them. 

\subsection*{Code availability}

The codes used in this work are available in the open GitHub repository \url{https://github.com/mridenti/COVID-DEMOGRAPHY}. We kindly ask that eventual users of the codes cite this article and \cite{ridenti2020} and the repository address on any publication that use them. 

\subsection*{Authors' contributions}

Marco A. Ridenti, Lara K. Teles and Vladimir K. Teles worked on the model formulation, implementation and interpretation of results. Vladimir K. Teles and Marco A. Ridenti did the data curation. Alexandre Maranh\~ao implemented the code that computes the basic reproduction number by the next generation method. 

\section{Introduction}
The outbreak of the COVID-19 disease, caused by the novel coronavirus SARS-CoV-2, which began in Wuhan, China, in late 2019, has spread worldwide, and has been officially declared a global pandemic \cite{WHO-pandemic}. Since then, it caused unprecedented public health interventions, including from social distancing, the closure of schools, home-office to full lockdowns procedures, which reduced the instantaneous reproduction number and help to contain the disease. 
Several standard mathematical epidemiological models were used and other new ones were developed to predict the incidence of the epidemics in a spatial population through time. A common model used, for example, by the economists is the traditional epidemiological model SIR (susceptible-infected-recovered),  as summarized by \cite{atkeson2020} and \cite{stock2020}. This model was used to evaluate the optimal social distancing policies \cite{AAL2020} or incorporating them into economic models endogenizing consumption and labor supply decisions (e.g., \cite{ERT2020b,ERT2020a,jones2020,krueger2020}). Another approach in the epidemiology literature has been to study the dynamics of the pandemic in extended versions of the basic SIR model, introducing exposed (E) people and incorporating age-structured dynamics based on contact matrices at work, school, family and social environments, as in \cite{ferguson2020,kissler2020projecting,prem2020effect}. 

\begin{figure}[!htb]
\centering
\includegraphics[width=1.0\linewidth]{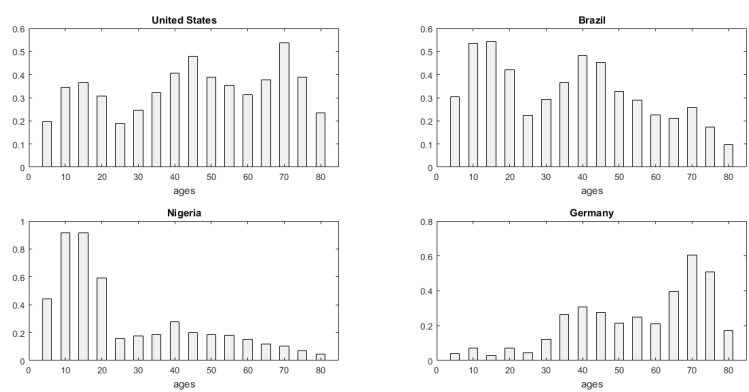}
\caption{Mean Age Contact Patterns of Population with an Age Higher than 65 years. Source: authors' calculations, original data from \cite{prem2017projecting}}
\label{fig1}
\end{figure}

\begin{figure}[!htb]
\centering
\includegraphics[height=9.5cm,width=1.0\linewidth]{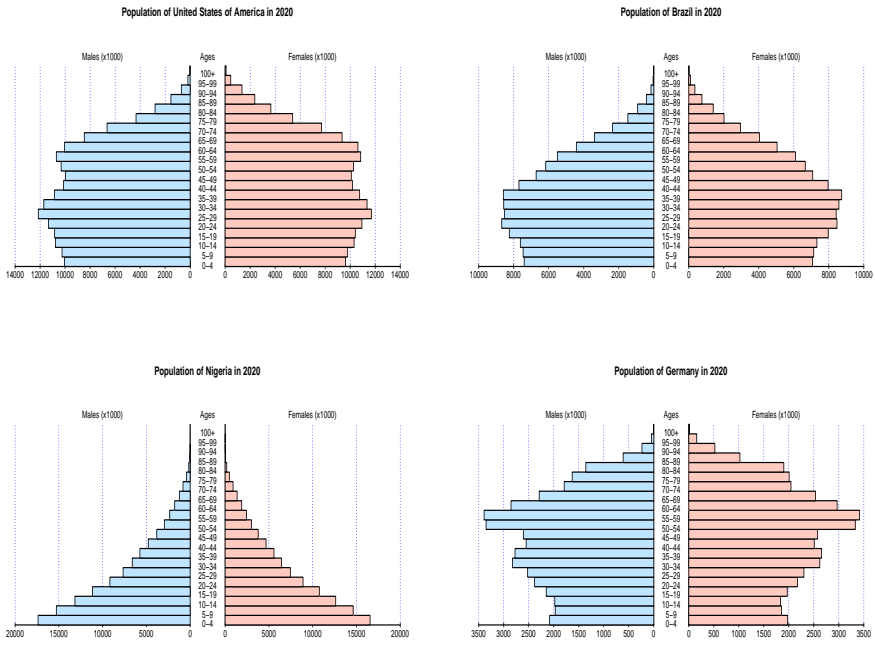}
\caption{Population Pyramids by Age and Gender. Source: World Population Prospects 2019. United Nations.}
\label{fig2}
\end{figure}

Among the several factors that affect the COVID-19 dissemination, there are the contact matrices, the demography, and the basic reproduction number $R0$.  At the same time, contact patterns and demography vary greatly among countries. To appreciate this, we can observe the examples presented in Figures \ref{fig1} and \ref{fig2}, where the contact patterns of the elderly population and population pyramids of the United States, Brazil, Nigeria and Germany are presented. Since person-to-person transmission is mostly driven by who interacts with whom, which can vary greatly by age, the prevalence, hospitalization and fatality rates of COVID-19 may be strongly dependent on those characteristics. 

However, as stressed by \cite{walker2020report}, the direction of the relationship between demography and the burden of COVID-19 is not clear. While developed countries present a much higher proportion of the population above 65, the age interval at particularly high risk of mortality from COVID-19, least developed countries present a much higher average household size of individuals aged 65 and over, increasing the potential spread specifically to this vulnerable age group. As documented by \cite{prem2017projecting}, demography and contact patterns walk together. These characteristics are a key context for COVID-19 transmission and fatality and may have ambiguous effects on the risk profile for COVID-19. Even more uncertain is how the effectiveness of social distancing measures may be compromised by demography and contact patterns. To illustrate, while Germans over 65 have more contact with others of the same age group and adults, in Nigeria, the elderly predominantly have contacts with children (Figure \ref{fig1}). Within these opposite social structures, school closure policies or restrictions on work may have distinct marginal effects in protecting the elderly from contamination.

Existing research has already emphasized the importance of demography, e.g., \cite{dowd2020demographic,kashnitsky2020covid,nepomuceno2020besides,sudharsanan2020contribution,balbo2020demography,esteve2020vulnerabilitat,goldstein2020demographic} and contact networks, e.g., \cite{bayer2020intergenerational,liu2020secondary,mossong2008social,balbo2020strength,karmakar2021association,papageorge2021socio} in explaining the spread of COVID-19. However, it is necessary to quantify its role in the effectiveness of intervention policies. 

Besides all the aspects mentioned above, the basic reproduction number, $R0$, which is an indication of the transmissibility of the virus, representing the average number of new infections generated by an infectious person in a fully susceptible population, should vary because neither the contact matrix nor the demographic profiles are the same for different countries. Therefore, it should be of great use and interest the development of an epidemiological model that takes into account all these characteristics together and assess the interaction effects between them. 

In this paper, we develop a mathematical augmented epidemiological model to investigate how epidemic dynamics and the effectiveness of isolation policies, such as school closures, work restrictions, community distancing and isolation of the elderly, depend on demographics and contact patterns. The proposed augmented epidemiological model is based in an age-structured SEAHIR model that divide the population into susceptible (S), exposed (E), infected and asymptomatic (A), hospitalized (H), infected and symptomatic (I), and recovered (R) individuals to simulate COVID-19 dissemination. Four fictitious representative countries were build fitting the contact patterns and demographic parameters to least developed, developing, developed and aged developed countries, setting the same population size for all of them and all other epidemiological parameters remaining equal. Applying the SEAHIR model to simulate COVID-19 dissemination in these four fictitious representative countries, we obtain a quantitative assessment of the marginal importance of contact patterns and demographics on the spread of COVID-19. We complete our simulations by building six scenarios with alternative combinations of four social distancing policies (school closures, work restrictions, isolation of the elderly and community distancing) and compare their effectiveness under different demographic and contact network patterns.

Therefore, this paper is related to both areas of study, developing and applying a general and detailed epidemiological model that incorporates hospitalized individuals, accounting for the demand for medical care due to COVID-19, and carefully observing how the risk profile for COVID-19 could be different in countries with alternative contact patterns and demographics and, consequently, how the tradeoffs related to social distancing policies vary among countries.

The paper is organized as follows: the next section describes the SEAHIR model. Section 3 presents the methodology and calibrates the model parameters. Section 4 discusses the results, and section 5 concludes the paper.

\section{The age-structured SEAHIR model}

The model that we propose to study the relationship between demography, contact patterns, and the effectiveness of social distancing policies adds asymptomatic and hospitalized groups to the model recently proposed by \cite{prem2020effect} being more appropriate to mimic the COVID-19 dissemination.

\begin{figure}[!htb]
\centering
\includegraphics[width=\linewidth]{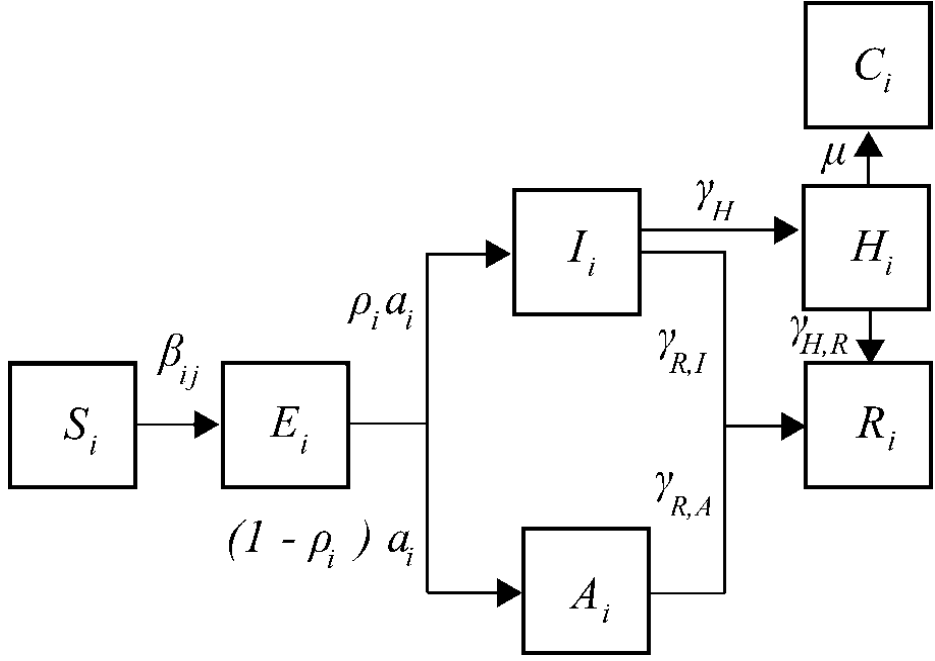}
\caption{SEAHIR model structure}
\label{model}
\end{figure}

Each period, a fraction of susceptible individuals ($ S $) is exposed to the virus. Such exposed individuals ($ E $) become infected and may become symptomatic ($ I $) or asymptomatic ($ A $). The transmission probability from exposed or asymptomatic individuals to susceptible individuals is not null but is smaller than the transmission probability from symptomatic individuals to susceptible individuals. While all asymptomatic patients recover ($ R $) over time, symptomatic patients may recover or be hospitalized ($H$), and the latter, in turn, may recover or die ($ C $). The total sum of living individuals is represented by the variable $ N $. (Figure \ref{model})

All parameters governing each transition between the health status of individuals are age specific. Thus, we consider age groups with intervals of 5 years, from 0 to 74 years of age, and an open interval $ \geq $ 75 years, comprising a total of 16 age groups. Therefore, the model is defined by a system of 96 differential equations composed of 16 age groups and 6 health statuses. In this age-structured model, we use subscribed indices ($i = 1$ to $16$) to indicate the age range of a dynamic variable. The equations of the SEAHIR model can be generalized and written as

\begin{eqnarray}\label{eq:model}
\frac{d}{dt}S_{i} & = & \delta_{1i}\Lambda_{i} N_{i} - \mu_{eq,i} S_{i} - \sum_{j} \beta_{ij} \frac{S_{i}}{N_{i}} I_{j} \nonumber \\
& & - \sum_{j} \alpha_{j} \beta_{ij} \frac{S_{i}}{N_i} A_{j} - \sum_{j} \xi_{j} \beta_{ij} \frac{S_{i}}{N_i} E_{j} \\
\frac{d}{dt}E_{i} & = & \sum_{j} \beta_{ij} \frac{S_{i}}{N_i} I_{j} + \sum_{j} \alpha_{j} \beta_{ij} \frac{S_{i}}{N_i} A_{j}
+ \sum_{j} \xi_{j} \beta_{ij} \frac{S_{i}}{N_i} E_{j} \nonumber \\ & & - \mu_{eq,i}E_{i} - a_{i}E_{i} \\
\frac{d}{dt}A_{i} & = & (1-\rho_{i}) a_{i} E_{i} - \mu_{eq,i}A_{i} - \gamma^{R,A}_{i} A_{i} \\
\frac{d}{dt}H_{i} & = & \gamma^{H}_{i} I_{i} -\gamma^{H,R}_{i} H_{i} - (\mu_{eq,i} + \mu_{cov,i}) H_{i} \\
\frac{d}{dt}I_{i} & = & \rho_{i} a_{i} E_{i} - \mu_{eq,i}I_{i} - \gamma^{H}_{i} I_{i} - \gamma^{R,I}_{i} I_{i} \\
\frac{d}{dt}R_{i} & = & \gamma^{R,I}_{i} I_{i} + \gamma^{R,A}_{i} A_{i} + \gamma^{H,R}_{i} H_{i} - \mu_{eq,i} R_{i} \label{eq:model_last} \\
\end{eqnarray}

The death rate coefficient under normal conditions is given by $ \mu_ {eq} $, and $ \mu_ {cov} $ is the COVID-19 age stratified death rate coefficient. The other parameters are $ \Lambda $, the population's typical birth rate coefficient; $ \beta_{ij} $, the contamination rate matrix; $ \gamma^ {R, I} $, the typical symptomatic recovery rate coefficient; $ \gamma^{R, A} $, the typical asymptomatic recovery rate coefficient; $\gamma^{H}$, the hospitalization rate coefficient; $\gamma^{H, R}$, the typical hospitalized recovery rate coefficient; $ a $, the exposed-to-infected conversion rate coefficient; $ \alpha $, a parameter that represents how much lower asymptomatic infectivity is than symptomatic infectivity; $\xi$, a parameter that represents how much lower presymptomatic infectivity is than symptomatic infectivity ($ \alpha $ and $ \xi $ must be a positive value less than or equal to 1); and $ \rho$, which is the probability that an exposed person will become symptomatic.

A set of key parameters in explaining how demographics and account contact patterns for the spread of the virus is defined in the $ \beta_{k,j} $ matrix, which should be interpreted as the rate of infection of susceptible individuals in the $i$ age range by infected individuals in the $j$ age range. In that sense, $ \beta_{k,j} $ is given by multiplying the probability of contagion or susceptibility $ p_i$ and the social contact matrix $ C_{k,j} $, which is described by the sum of the interaction matrices of the contacts at home, $C_{kj}^{H}$, work $ C_{kj}^{W} $, school $ C_{ kj}^{S} $ and other sources $ C_{kj}^{O}$

\begin{equation}\label{eq:total_contact}
C_{kj} = T\left( C_{kj}^{H} + C_{kj}^{W} + C_{kj}^{S} + C_{kj}^{O} \right) \,\, .
\end{equation}
where $T$ is a transformation of the contact matrices describing the effect of social distancing measures.

\section{Methodology}

\subsection{Calibration}

To perform the simulations based on the SEAHIR model, we calibrate the parameters that govern the differential equation system based on up-to-date knowledge about COVID-19.

The age-specific hospitalization parameters were obtained from \cite{verity2020estimates} according to the statistics of Chinese cases corrected for underreporting and demographic profile. In this model, the probability of a case being clinical/symptomatic ($\rho_{i}$) and the susceptibility $p_i$ are age dependent. The values of $\rho_{i}$ and $p_i$ were extracted from the work of Davies \emph{et al} \cite{davies2020age}, who estimated these parameters by fitting an age-structured mathematical model to epidemic data from China, Italy, Japan, Singapore, Canada and South Korea. 

With respect to time span parameters, we considered the average infection time ($d_{I}$) 5 (3 - 7) days \cite{woelfel2020clinical}, the average incubation time ($d_{L}$) 7 (4.6 - 9) days \cite{lauer2020incubation}, the average time from infection to death ($d_{C}$) 15.0 (12.5 - 17.8) days \cite{sun2020understanding,linton2020incubation}, the time from onset of infection to hospitalization ($d_{H}$) 6 (3-7) days \cite{zhou2020clinical,guan2020clinical}, and the typical time in hospitalization until discharge ($d_{A}$) 8 (4-15) days \cite{zhou2020clinical}. Using these time span parameters, we obtained the recovery rate coefficient of symptomatic infection ($\gamma_{R,I} \sim 1 - e^{-(1/d_{I})}$), the recovery rate coefficient of asymptomatic infection ($\gamma_{R,A} \sim 1 - e^{-(1/d_{I})}$), the hospitalized recovery rate coefficient ($\gamma_{H,R} \sim 1 - e^{-(1/d_{A})}$), and the incubation period conversion rate coefficient ($ a \sim 1 - e^{-(1/d_{L})}$).

The hospitalization rate coefficient of the infected is given by

\begin{equation}\label{rel_gammas}
\gamma_{H} = \gamma_{R,I}\phi/(1-\phi)
\end{equation}
where $\phi$ is the percentage of infected persons requiring ICU admission \cite{verity2020estimates}. 

The parameter $\mu_{cov}$ is conditioned by the prescribed infection fatality rate, $ T_{L, C} $, and is estimated by the equation

\begin{equation} \label{rel_mu}
\mu_{cov} \sim T_{L,C} \frac{1 - \phi}{\phi} ( \gamma_{H} + \gamma_{RI}) \,\, .
\end{equation}
These relations between the parameters must be satisfied so that the model results remain always consistent with the prescribed infection fatality rate ($ T_{L, C} $) and the percentage of infected persons requiring ICU ( $\phi$). In Appendix \ref{appendix}, these expressions are derived from the model system of equations (Eqs. \ref{eq:model} to \ref{eq:model_last}) . We consider that the infection fatality rate ($ T_{L, C} $) and the percentage of infected persons requiring ICU ($\phi$) are distributed among age groups according to Wuhan statistics estimated by \cite{verity2020estimates}.

The correction factor for the subclinical infection coefficient, $\alpha$, was considered to be 0.75 and the correction factor for the presymptomatic infection coefficient, $\xi$, was set equal to 0.5 following the CDC recommended values \cite{CDCrecommend}.

\subsection{Demography and Contact Patterns}\label{contact_R0}

To measure the importance of demographics and contact patterns for the dissemination of COVID-19, we built four representative fictitious countries whose demographic and contact patterns refer to stages of demographic transition and applied the SEAHIR model simulations to them.

We keep all general parameters constant, as calibrated in the previous session, and assign the four representative fictitious countries the same population size (10 million inhabitants), since distinct population sizes would affect the dynamics of the spread of the disease. The initial condition is always the same for all simulations and is set as follows: the infected population is set to one, the susceptible population is set to the total number of individuals (population size) and all the other dynamical variables are set to zero.

In addition, we assumed that the average probability of getting infected after one contact ($p_i$) depends only on the age of the susceptible individuals. We considered that this probability is the same for all countries. We note that the average number of contacts between individuals depends on the country, since it reflects the specific social patterns of the country, but the susceptibility depends only on the nature of the disease and the typical human immunological response to it. The susceptibility may vary due to differences on virus strains or genetic differences among different individuals, but these effects were not taken into account in our simulation. By fixing the susceptibility per age group for all countries and all the model parameters, we may compute the expected variations on the epidemiological dynamics that are only due to differences of contact patters and demographic profiles. 

As a consequence of fixing the susceptibility per age group, the basic reproduction number $R0$ should vary because neither the contact matrix nor the demographic profiles are the same for different countries. We could eventually adjust the susceptibility in order to get the same $R0$ value for all countries, but this would be unrealistic, since the basic reproduction number should naturally vary both as a function of the intrinsic disease biological characteristics and the population behavior. Therefore, we shall first determine the basic reproduction number $R0$ without intervention for each country. This would give us information of the disease dynamics which are only due to social and demographic factors. Next, we shall compute how different interventions affect the basic reproduction number for different countries. 

The $R0$ was computed exactly using the next-generation method \cite{diekmann1990definition,van2017reproduction}. In this method, the right-hand side of the system of equations (Eqs. \ref{eq:model} to \ref{eq:model_last}) is split in a creation term $\mathcal{F}_i$, representing the rate of new infections in compartment $i$, and a loss term $\mathcal{V}_i$, representing the rate of transition of infected to other infection compartments. By defining the matrices $F_{i,j}=\partial \mathcal{F}_i/\partial x_j$ and $V_{ij}=\partial \mathcal{V}_i/\partial x_j$, where $x_{j}$ is any compartment with infected individuals, the basic reproduction number is computed as the spectral radius of the matrix $\mathbf{FV}^{-1}$ (see Appendix \ref{appendix2}). We could eventually derive a complicated analytical expression for the $R0$ value, but as the model contains too many compartments and parameters, we preferred to build a computational routine to compute the $R0$ values from any given set of model parameters and contact matrices.

We simulated the dissemination of COVID-19 in four alternative demographic and contact patterns. As highlighted by \cite{prem2017projecting}, there is a clear relationship between demographics and the contact patterns of people and generations. The size of families, the number of students in school classrooms, the number of people living in the same house and people's occupation are determinants of the contact matrix and are determined by demographics. Thus, when classifying countries into stages of demographic transition, we automatically define their patterns of interpersonal and intergenerational contact.

In general, we can establish four patterns to represent countries according to their stage of demographic transition in the current world:

\textbf{Least Developed Countries (LDCs)}: have a high fertility rate and rapid population growth, with the demographic pyramid having a triangular shape;

\textbf{Developing Countries (DGCs)}: women have an increase in their status, they have access to contraception, they have a reduction in the fertility rate, and the rate of population growth begins to be reduced. The age pyramid has a transition format;

\textbf{Developed Countries (DCs)}: birth and death rates are low, stabilizing the size of the population;

\textbf{Aged Developed Countries (ADCs)}: the fertility rate has fallen well below that of fertility, where the elderly population now represents a considerable share of the population.

The contact matrices $C_{ij}$ and demographic parameters that we will use to represent each of these demographic stages came from Nigeria (Least Developed Countries Stage), Brazil (Developing Countries Stage), United States (Developed Countries Stage), and Germany (Aged Developed Countries)\footnote{Since all other parameter do not follow the values of these countries, our simulations should not be interpreted as forecasts for them.}. The data sources are the United Nations' World Population prospects and \cite{prem2017projecting}.

\subsection{Policy Interventions}

To assess the effectiveness of intervention policies in different environments for contact patterns and demography, we simulated the implementation of six combinations of four alternative types of intervention (school closures, community distancing, restrictions on work and isolation of elderly individuals) for one year. In all simulations, the intervention starts after the time when around 100 individuals are currently infected. 

\subsubsection{Scenario 1: No intervention}

The total contact matrix is given by Eq. \ref{eq:total_contact}.

\subsubsection{Scenario 2: School closure, community distancing, work restrictions and isolation of the elderly }

Through scenarios 1 to 6, this is the strictest one. In this scenario, the school matrix is zeroed, and a decrease in youth contacts in the community is applied to the $ C_{kj}^{O}$ matrix. A decrease in work contacts is also considered. 

At the same time, since all individuals spend more time at home, we consider an increase in youth contact (0 to 20 years) and a minor increase in other age groups in the home matrix. To restrict the contacts of the elderly to the home, a reciprocal transformation should be performed on the matrices to reduce only the rows and columns corresponding to the older population range:

\begin{equation}
C_{kj} = A_{kk} C_{kj}^{H} + \zeta_{t} (\mathbf{\tilde{I}} \cdot \mathbf{C}^{W} \cdot \mathbf{\tilde{I}} )_{kj} + 0 \cdot ( \mathbf{\tilde{I}} \cdot \mathbf{C}^{S} \cdot \mathbf{\tilde{I}})_{kj} + B_{kk} ( \mathbf{\tilde{I}} \cdot \mathbf{C}^{O} \cdot \mathbf{\tilde{I}})_{kj}
\end{equation}
where $ \mathbf{A}$ is a diagonal matrix, \emph{e.g.}, $A_{ii} = 1.5$ for $ i = 1 $ to $ 4 $ and $ A_{ii} = 1.1 $ for $ i> 4 $, in the case of a $ 50\% $ increase for young people and $ 10 \% $ increase for adults and the elderly; $ \mathbf{B} $ is a diagonal matrix, \emph{e.g.}, $B_{ii} = 0.4 $ for $ i = 1 $ to $ 4 $, and $ B_{ii} = 0.6 $ for $ i> 4 $ ($ 60 \% $ decrease in youth community contacts and $ 40 \% $ decrease for other age groups); and $\zeta_{t}$ is a scalar, with $\zeta_{t} = 0.5 $ for all $i$, in the case of a 50 $\%$ decrease in work contacts.

Some models consider total isolation of the elderly, \emph{i.e.}, no contacts with adults and younger individuals not only outside home, but also no contacts at home. We consider this supposition too idealistic and unfeasible, since old people require the aid of younger people at home. For this reason, we preferred to keep unchanged the contact matrix of old people at home. 

\subsubsection{Scenario 3: School closure, community distancing and isolation of the elderly }

In this intervention scenario, the contacts of the elderly are restricted to the home. We consider a prescribed increase in youth contact (0 to 20 years) and a minor increase in other age groups in the home matrix. The school matrix is zeroed. A prescribed decrease in contacts in the community is applied to the $ C_{kj}^{O}$ matrix.

\begin{equation}
C_{kj} = A_{kk} C_{kj}^{H} + (\mathbf{\tilde{I}} \cdot \mathbf{C}^{W} \cdot \mathbf{\tilde{I}} )_{kj} + 0 \cdot ( \mathbf{\tilde{I}} \cdot \mathbf{C}^{S} \cdot \mathbf{\tilde{I}})_{kj} + B_{kk} ( \mathbf{\tilde{I}} \cdot \mathbf{C}^{O} \cdot \mathbf{\tilde{I}})_{kj}
\end{equation}
where $ \mathbf{A}$ is a diagonal matrix, \emph{e.g.}, $A_{ii} = 1.5$ for $ i = 1 $ to $ 4 $ and $ A_{ii} = 1.1 $ for $ i> 4 $, in the case of a $ 50\% $ increase for young people and $ 10 \% $ increase for adults and the elderly; and $ \mathbf{B} $ is a diagonal matrix, \emph{e.g.}, $B_{ii} = 0.4 $ for $ i = 1 $ to $ 4 $, and $ B_{ii} = 0.6 $ for $ i> 4 $ ($ 60 \% $ decrease in youth community contacts and $ 40 \% $ decrease for other age groups).

\subsubsection{Scenario 4: School closure and community distancing}

In this case, we consider a prescribed increase in youth contact (0 to 20 years) and a minor increase in other age groups in the home matrix. The school matrix is zeroed. A prescribed decrease in contacts in the community is applied to the $ C_{kj}^{O}$ matrix,

\begin{equation}
C_{kj} = A_{kk} C_{kj}^{H} + C_{kj}^{W} + 0\cdot C_{kj}^{S} + B_{kk} C_{kj}^{O} \,\, ,
\end{equation}
where $ \mathbf{A}$ is a diagonal matrix, \emph{e.g.}, $A_{ii} = 1.5$ for $ i = 1 $ to $ 4 $ and $ A_{ii} = 1.1 $ for $ i> 4 $, in the case of a $ 50\% $ increase for young people and $ 10 \% $ increase for adults and the elderly, and $ \mathbf{B} $ is a diagonal matrix, \emph{e.g.}, $B_{ii} = 0.4 $ for $ i = 1 $ to $ 4 $ and $ B_{ii} = 0.6 $ for $ i> 4 $ ($ 60 \% $ decrease in youth community contacts and $ 40 \% $ decrease for other age groups).

\subsubsection{Scenario 5: Isolation of the elderly and community distancing}

In this intervention scenario, the contacts of the elderly are restricted to the home. There is no change in the home matrix, since now the schools are open. A prescribed decrease in contacts in the community is applied to the $ C_{kj}^{O}$ matrix.

\begin{equation}
C_{kj} = C_{kj}^{H} + (\mathbf{\tilde{I}} \cdot \mathbf{C}^{W} \cdot \mathbf{\tilde{I}} )_{kj} + ( \mathbf{\tilde{I}} \cdot \mathbf{C}^{S} \cdot \mathbf{\tilde{I}})_{kj} + B_{kk} ( \mathbf{\tilde{I}} \cdot \mathbf{C}^{O} \cdot \mathbf{\tilde{I}})_{kj}
\end{equation}
where $ \mathbf{B} $ is a diagonal matrix, \emph{e.g.}, $B_{ii} = 0.4 $ for $ i = 1 $ to $ 4 $, and $ B_{ii} = 0.6 $ for $ i> 4 $ ($ 60 \% $ decrease in youth community contacts and $ 40 \% $ decrease for other age groups).

\subsubsection{Scenario 6: Community distancing}

A prescribed decrease in contacts in the community is applied to the $ C_{kj}^{O}$ matrix, but there is no change in the home matrix, since now the schools are opened, and we have

\begin{equation}
C_{kj} = C_{kj}^{H} + C_{kj}^{W} + C_{kj}^{S} + B_{kk} C_{kj}^{O} \,\, ,
\end{equation}
where $ \mathbf{B} $ is a diagonal matrix, \emph{e.g.}, $B_{ii} = 0.4 $ for $ i = 1 $ to $ 4 $, and $ B_{ii} = 0.6 $ for $ i> 4 $ ($ 60 \% $ decrease in youth community contacts and $ 40 \% $ decrease for other age groups).

\subsubsection{Scenario SL: Strong Isolation}

This is the closest scenario to a full lockdown, used as a reference scenario for which all countries would experience a basic reproduction number $R0$ lower than one. This case is similar to scenario 2, but the work contacts are reduced by $70\%$ and the community contacts are reduced by $90\%$ for all age groups. This scenario was only used to serve as an example of social isolation levels that could effectively stop the epidemic. 

\section{Results}

\begin{figure}[!htb]
\centering
\includegraphics[width=1.0\linewidth]{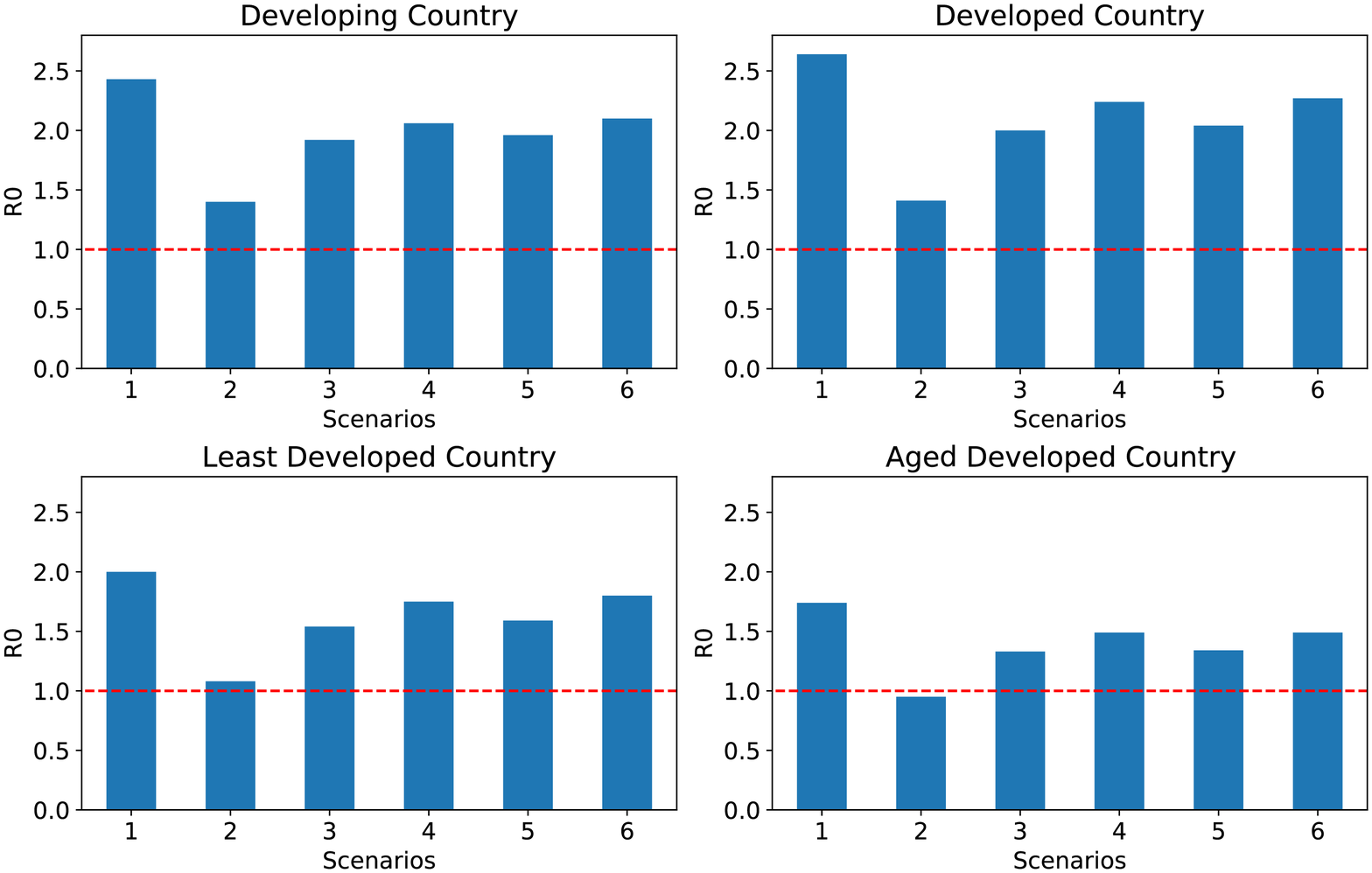}
\caption{Basic reproduction number $R0$. \footnotesize{ \textbf{Scenario 1}: No intervention; \textbf{Scenario 2}: School closures + isolation of the elderly + community distancing + work restrictions; \textbf{Scenario 3}: School closures + isolation of the elderly + community distancing; \textbf{Scenario 4}: School closures + community distancing; \textbf{Scenario 5}: Isolation of the elderly + community distancing; \textbf{Scenario 6}: Only community distancing during the full year;}}

\label{r0}
\end{figure}

When simulating the SEAHIR model for the four representative fictitious countries and the six policy intervention scenarios, we evaluated the results based on four indicators: basic reproduction number ($R0$), prevalence after one year, demand for hospitalizations at the maximum point, and fatality rate (deaths/100 k) after one year.

The computed basic reproduction numbers $R0$ for each scenario and country are shown in Figure \ref{r0}. The results show the differences between $R0$ values among different countries and the quantitative impact of intervention policies on their values. The $R0$ values without intervention are 2.43 (developing country), 2.64 (developed country), 2.00 (least developed country), 1.74 (aged developed country). These $R0$ values are consistent with literature values estimated in the early epidemic breakdown in China, as reported by a review by \cite{liu2020reproductive} yielding expectation values around 2-3. China can be classified among the group of developing or developed countries, so the results are consistent with these literature estimates. The consistency of the computed reproduction number warrants that the present model should give reasonable estimates of other relevant quantities, such as fatality and prevalence. High accuracy should not be expected, however, as the model contain some uncertain or simplifying assumptions and some of its parameters are not known accurately. All the results and conclusions of this simulation are valid as far as the basic assumptions of the model hold. Despite these limitations, simulation results may be insightful as they allow us to draw the connection between different demographic and social characteristics and their ultimate impact on the epidemic behavior. 

As we have already discussed earlier, the $R0$ is not a constant parameter because it depends on the typical behavior of the population where the epidemic spreads. Curiously, our results show that the least developed country and the aged developed country have the lowest $R0$ values. This result has already been observed in a previous work where age dependent susceptibilities were also considered \cite{hilton2020estimation}. Individuals in aged developed countries usually have much lower social contact rates both at home and outside home than least developed countries. We computed the average number of contacts an individual has during the time she/he is infectious, and we find that in the least developed country this number is 24 while in the the aged developed country this number is 4. It is not hard therefore to understand why the aged developed country should have lower reproduction numbers given the assumptions of the present model. The least developed country has a low $R0$ value for a different reason: it has an expansive population pyramid while the susceptibility among younger people is much lower. These factors compensate its much higher social contact rates, since contacts occur mainly between young people. The highest values of $R0$ were observed for the developed and developing countries, because they have higher social contact rates than aged developed country while having a significant ratio of its population in the older age groups. 

The strictest intervention policy (scenario 2) reduced by around $45\%$ the basic reproduction number for all countries, without pronounced differences between them ($42\%$ DGC, $45\%$ ADC, $46\%$ LDC, $46\%$ DC ). In this scenario, the $R0$ value in the aged developed country dropped to less than one, and the $R0$ value in the least developed was higher but very close to one. By allowing work contacts (scenario 3), the decrease was around $20\%$ ($21\%$ DGC, $24\%$ ADC, $23\%$ LDC, $24\%$ DC). By lifting the isolation of elderly people (scenario 4), the decrease was $15\%$ for the DGC, $14\%$ for the ADC, $12\%$ for the LDC, and $15\%$ for the DC. In this scenario the LDC had the lowest $R0$ decrease, which can be readily attributed to its relatively smaller old population. In scenario 5, where the elderly are isolated, but schools are opened, the decrease was $19\%$ for the DGC, $23\%$ for the ADC, $20\%$ for the LDC, and $23\%$ for the DC. The school opening (scenario 5) in the aged developed country had very little impact in the $R0$ value, with a decrease of around $2\%$ in relation to scenario 3, but the impact was higher on the least developed and developing countries, with decreases of $11\%$ and $8\%$ in relation to scenario 3. Finally, in scenario 6, where only community contacts are restricted, the decrease in $R0$ was $14\%$ for the DGC, $14\%$ for the ADC, $10\%$ for the LDC, and $14\%$ for the DC. For the ADC, there was virtually no difference in terms of $R0$ between scenario 4 (school closing and community distancing) and scenario 6 (community distancing), so school closing had low impact on the epidemics for this country. School closing had a stronger effect on the LDC and the DGC, where the relative $R0$ decrease from scenario 4 (school closing and community distancing) to scenario 6 (community distancing) was $20\%$ and $11\%$, respectively.

We may conclude from this analysis that restrictions in work and community contacts had higher impact on the decrease of $R0$ values, with elderly isolation coming next and school closing at last. The relative decreases in $R0$ for each scenario are similar, with some small differences. Therefore, countries having higher basic reproduction numbers, such as developing and developed countries, need to adopt more restrictive measures - or alternative non-pharmaceutical intervention strategies - in order to reduce the reproduction number to lower levels. We also computed the expected $R0$ in the case of strong isolation (SL scenario), resulting in 0.87 for the DGC, 0.65 for the ADG, 0.70 for the LDC and 0.90 for the DC. In this scenario, all $R0$ values were decreased to levels below one, but the lowest levels were those from the least developed and aged developed countries, because their basic reproduction numbers without intervention were lower.



\begin{figure}[!htb]
\centering
\includegraphics[width=1.0\linewidth]{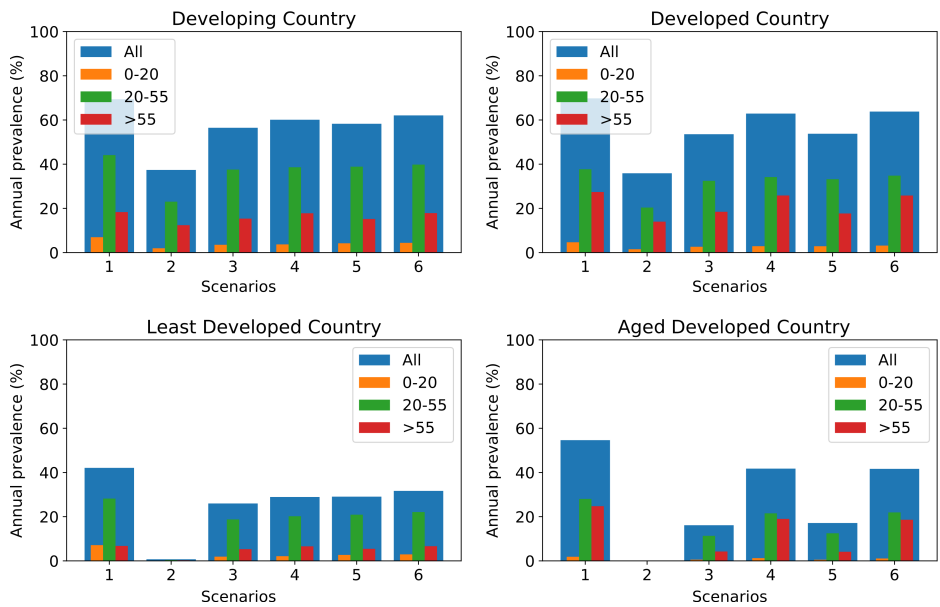}
\caption{Prevalence after one year (\%) \footnotesize{ \textbf{Scenario 1}: No intervention; \textbf{Scenario 2}: School closures + isolation of the elderly + community distancing + work restrictions; \textbf{Scenario 3}: School closures + isolation of the elderly + community distancing; \textbf{Scenario 4}: School closures + community distancing; \textbf{Scenario 5}: Isolation of the elderly + community distancing; \textbf{Scenario 6}: Just community distancing during one year;}}
\label{prevalence}
\end{figure}

\newpage

The simulated results for prevalence are presented in Figure \ref{prevalence}. The aged developed country presented the higher prevalence in the scenario with no intervention (69.8$\%$), and the least developed country had the lowest prevalence (42.1$\%$), since it has a higher proportion of young people who present a lower symptomatic infection rate and susceptibility. After the strictest scenario (scenario 2), the developing country had the higher prevalence (37.4$\%$), while the aged developed country and the least developed country had almost zero prevalence due to the very low $R0$ value. Therefore, this social distancing policy had a high relative impact in aged and least developed countries, since the prevalence decreased from 54.2$\%$ to 0.1$\%$ in the former and decreased from 42.1$\%$ to 0.7$\%$ in the latter. In the developing and least developing countries the decrease in the prevalence was not as marked as in the previous cases, decreasing from 69.8$\%$ to 37.4$\%$ in the DGC, and 69.8$\%$ to 35.9$\%$ in the DC. This can be explained by the higher $R0$ value of the epidemic in these countries. Also, because the in-house contacts of the elderly is not restricted in this model, their corresponding compartment would still be susceptible to infection and fatalities. 

By comparing now all the intervention scenarios, it can be readily seen that the work contact reduction was the intervention giving the largest contribution to the decrease of prevalence and $R0$, which has already been noted earlier. Although the prevalence reduction in the developed and developing countries were not as large as in the other cases after intervention (scenario 2), the work contact reduction is responsible for more than 50\% of the effectiveness of the intervention. Let's now analyze the effectiveness of the other interventions. 

The effectiveness of a simple community distancing policy (scenario 6) was very similar between the developing and developed countries, with a relative decrease in prevalence ranging from 8.6 \% (developed country) to 10.5 \% (developed country). The relative decrease for the aged developed country and least developed country were higher, 24.7 \% in the former and 23.8 \% in the latter. This difference may be again explained by the lower final value of $R0$ after the simple community distancing policy. The scenario 6 is important as it allow us to study separately the effectiveness of the school closure (scenario 4) and the isolation of the elderly (scenario 5). Hereafter, in the following discussion of the prevalence, all the relative variations were computed taking as reference the prevalence value from scenario 6.

Scenarios 4 (school closure) and 5 (isolation of the elderly) presented strongly different effectiveness, in comparison with scenario 6, between countries due to demography. The isolation of the elderly reduces the overall prevalence by 58.8\% in aged developed countries but only 6.1\% in the developing country, 8.2 \% in the least developed country and 15.7\% in the developed country. 
On the other hand, school closure reduces the overall prevalence by 8.7\% in the least developed country and 3.1\% in the developing country, but only 1.4\% in the developed country. In the aged developed country there is a small detrimental effect of -0.2\%. Here, we can observe that even with the limited effectiveness of school closure, it presents a higher overall reduction in prevalence than the isolation of the elderly in the least developed country. School closure was 7.4\% more effective than isolation of the elderly in the least developed country, while isolation of the elderly was 11.2 times more effective than school closure in the  developed country and twice more effective in the developing country. We conclude here that isolation of the elderly had more impact when compared with school closure, except in the case of the least developed country. 

\begin{figure}[!htb]
\centering
\includegraphics[width=1.0\linewidth]{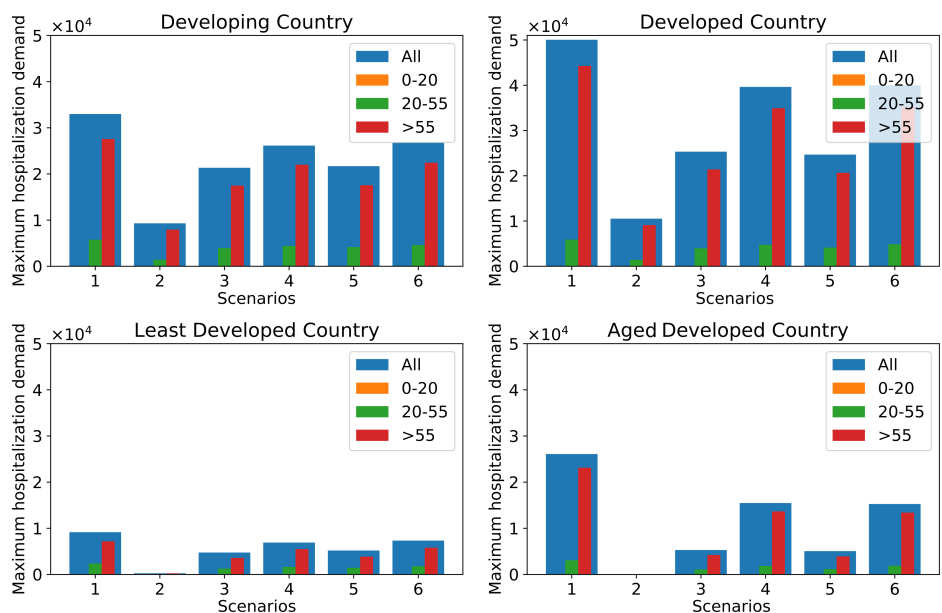}
\caption{Maximum Demand for Hospitalization. \footnotesize{ \textbf{Scenario 1}: No intervention; \textbf{Scenario 2}: School closures + isolation of the elderly + community distancing + work restrictions; \textbf{Scenario 3}: School closures + isolation of the elderly + community distancing; \textbf{Scenario 4}: School closures + community distancing; \textbf{Scenario 5}: Isolation of the elderly + community distancing; \textbf{Scenario 6}: Only community distancing during the full year;}}
\label{hospitalization}
\end{figure}

The simulation results for the maximum demand for hospitalizations are presented in Figure \ref{hospitalization}. In the case of no intervention, the number of hospitalizations was 5.4 times higher in the developed countries than in the least developed country. The highest value was observed in the developed country, since it has the highest $R0$ value and a relatively high old population. The peak number of hospitalizations was 1.3 higher in the developing country than in the aged developed country, despite the latter having a much higher relative old population. In this case, the lower $R0$ value of the aged developed country compensates the increased risk of hospitalization due to its higher average age. 

The maximum demand for hospitalization is virtually reduced to zero after the strictest intervention (scenario 2) in the aged and least developed countries due to the large decrease in the $R0$ value. In the developing country the maximum demand for hospitalization was reduced by 71.8 $\%$ and in the developed country it was reduced by 78.9 $\%$. The relative reduction on the maximum hospitalization demand is the highest among the four quantities in analysis (the other are $R0$, prevalence and fatality), which shows that intervention policies can be quite effective to alleviate the epidemic burden on the healthcare system. 

Community distancing (scenario 6) reduced maximum demand for hospitalization with a similar effectiveness between the developing and developed countries (10.5\% and 8.6\%, respectively), and a similar effectiveness between the aged and least developed countries (23.7 \% and 24.7 \%). Scenarios 4 (school closure) and 5 (isolation of the elderly) in comparison with scenario 6 present different patterns between aged developed countries and least developed countries, following the same direction of prevalence. The qualitative behavior is essentially the same, except for the fact that the isolation of the elderly is always the intervention policy with the highest impact. 

\begin{figure}[!htb]
\centering
\includegraphics[width=1.0\linewidth]{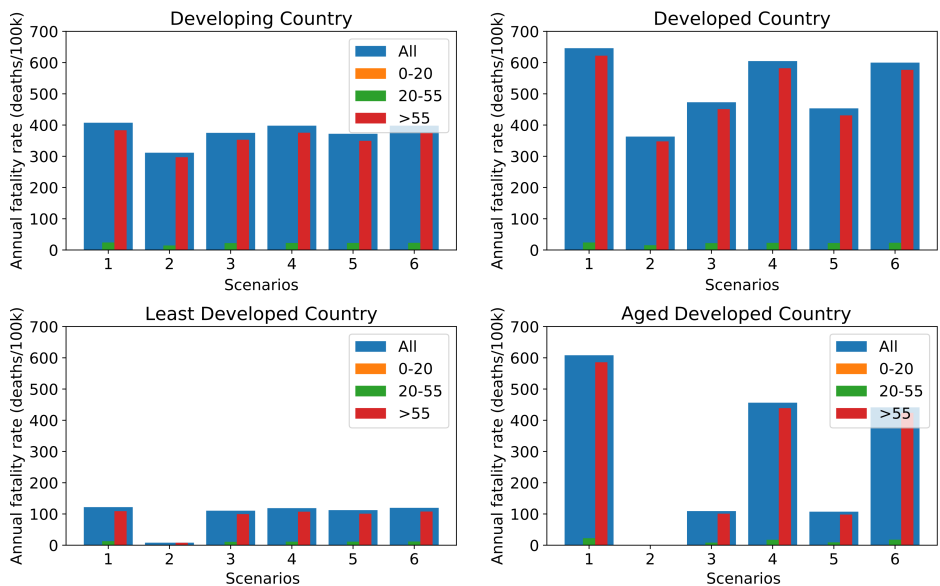}
\caption{Annual Fatality Rate (deaths/100k). \footnotesize{ \textbf{Scenario 1}: No intervention; \textbf{Scenario 2}: School closures + isolation of the elderly + community distancing + work restrictions; \textbf{Scenario 3}: School closures + isolation of the elderly + community distancing; \textbf{Scenario 4}: School closures + community distancing; \textbf{Scenario 5}: Isolation of the elderly + community distancing; \textbf{Scenario 6}: Only community distancing during the full year;}}

\label{fatality}
\end{figure}

Finally, simulation results for the annual fatality rate (deaths/100k) are presented in Figure \ref{fatality}. There are substantial differences in fatality rates between countries due to demography. In the scenario without interventions (scenario 1), the aged developed and the developed countries presented 608 and 646 deaths per 100,000 people after one year, respectively, while the developing and least developed countries presented 407 and 121 deaths per 100,000 people after one year. Although the aged developed country had the lowest $R0$ value, it was the second country with the highest fatality rate. This may be explained by the high prevalence among old people, which is a direct consequence of a demographic profile with the highest ratio of old individuals. The developed country had the highest fatality rate, which is a consequence of both its relatively high fraction of old people in the total population and the high $R0$ value. The relatively small value of the fatality rate in the least developed country is simply explained by the very small fraction of the population in the group of individuals being 55 years-old or more.

The fatality rates computed in the simulations are higher than the ones observed in the year 2020. For instance, by consulting the data of fatalities due to COVID-19 in these countries in 2020 \cite{covid_dataset}, we found that Brazil had 134 deaths/100k , Germany had 86 deaths/100k, Nigeria had 1 deaths/100k and US 155 deaths/100k. The real fatality rates reflect the actual intervention policies and many other factors which were not considered in the model, such as the contact pattern heterogeneity in big countries, the differences in susceptibilities due to different kinds of work and community interactions among individuals, the voluntary reduction in contacts, the use of masks and other personal hygiene measures that reduces the probability of infection. Also, the infected and asymptomatic did not have their contacts decreased in this model, when in reality they are often quarantined voluntarily or mandatorily. Despite that, the model explains the global trend, with the developed and developing country having the highest fatalities rates, and the least developed country having by far the lowest fatality rate. The model is not supposed to be accurate, but it predicts that the epidemic should be more difficult to control in a developing and developed country, due to their contact pattern and age profile. It also predicts a much lower impact in terms of fatality rates in the least developed countries. These features were indeed observed in the epidemic data from these countries. 

The strictest intervention policy (scenario 2) reduced by 23.5\% the fatality rate in the developing country and by 43.8\% the fatality rate in the developed country. The fatality rates in the least and the aged developed countries were reduced by almost 100 \% due to the reduction of the $R0$ value to a value near one. In the developing and least developed countries the work restrictions had the greatest impact in the reduction of fatality rates, responding for 89.9 \% and 66.2 \% of the decrease, respectively, while in the aged developed and developed countries the isolation of the elderly had the greatest impact, responding for 55.0 \% and 51.7\% of the decrease, respectively. The school closure had a small detrimental effect in the developed and age developed countries, for the same reasons discussed earlier, but it had a small beneficial effect in the least developed and developing countries. 

A remarkable result was that all intervention policies had a relatively small impact in the fatality rate in the developing country. The reason for that lies in the assumption of our model that the isolation of the elderly does not decrease the home contacts. In developing countries, grandparents often leave with their descendants (children and grandchildren) by need and not by option. If somehow the elderly could also live isolated, with no contacts but the essential, then this figure could drastically change. The relative decrease in fatality rate due to the isolation of the elderly in the developed and the least developed country is much higher because home contact rates between the young and the elderly in these countries are not as high as in the developing and least developed countries.

\section{Conclusions}

Since the outbreak of the SARS-CoV-2 (COVID-19) pandemic, countries have sought to contain the spread of the virus and protect the lives of their inhabitants. The most widely used measures have been the adoption of social isolation policies, with tradeoffs that appear to vary substantially between countries. This article sought to investigate the importance of contact patterns and demography in the spread and fatality of COVID-19 and, perhaps more importantly, how these factors influence the effectiveness of social distancing measures. By accounting for the magnitude of such effects, we can shed light on the size of the tradeoffs involved in the decision to implement such policies on a case-by-case basis.

To this end, we develop and implement an age-structured SEAHIR (susceptible - exposed - asymptomatic - hospitalized - infected (and symptomatic) - recovered) augmented epidemiological model to conduct simulations of six combinations of four types of isolation policies (work restrictions, isolation of the elderly, community distancing and school closures) to four representative fictitious countries generated over alternative demographic transition stage patterns (aged developed, developed, developing and least developed countries). The computer code used in this work to solve numerically the SEAHIR differential equations are available online on GitHub and it can be eventually adapted to other applications such as data fitting, specific demographic studies and integration into economic models.      

The results indicate that the burden of COVID-19 varies significantly between different demographic and contact patterns. The basic reproduction number $R0$ sets how fast the pandemic spreads and determines the prevalence of the disease, but it does not depend only on the intrinsic behavior of the virus and the human immunological response; our results showed that both the age profile and the contact patterns control how the epidemic evolves. An aged country could be expected to be among the most vulnerable to COVID-19 due to its large old population, but the results showed that when aged countries have very low contact rate between individuals, then their age vulnerability may be compensated. Indeed, our results showed that the epidemic in the aged developed and least developed countries had the lowest $R0$ values. It was shown that the intervention policies produced similar relative decreases in the $R0$ values in all the countries, but the values after intervention were lower where the ``no intervention'' $R0$ values were also lower, as in the aged developed and least developed countries. For instance, the strictest intervention reduced the basic reproduction number to a value lower or very near one in the aged developed and least developed countries, but not in the developing and developed countries. This suggests that some countries -- such as the developing and developed countries -- may have to increase the intervention strength, or at least search for alternative strategies.

 Here we highlight that what effectively reduces the basic reproduction number is either the decrease of the susceptibility or the decrease in contacts. For instance, individuals may be educated to learn how to reduce the probability of being infected, by means such as the use of masks, the practice of personal hygiene protocols etc., or individuals may be oriented to decrease as much as possible the frequency of close contacts with people outside their immediate family circle. Additional public health policies, not considered in this model, such as mass testing and contact tracing, could certainly help to mitigate the epidemic. 

Of all intervention policies, the $50\%$ reduction in work contacts was the one which produced the highest decrease in the $R0$ value, prevalence, maximum hospitalization demand and fatality rate, with few exceptions. Only in the case of the developed and aged developed countries, the elderly isolation produced a higher reduction in the fatality rate than the work restrictions. The community distancing was as important as the isolation of the elderly to reduce the computed quantities, and its impact was usually higher, except in the case of the aged and developed countries. The school closure was the intervention policy with the lowest relative impact on the reduction of the computed quantities, except in one case, in the least developed country, where school isolation had more impact in the prevalence reduction than the elderly isolation. We may conclude that in general the work reduction and the community isolation are effective interventions in all the modeled countries. The elderly isolation had greater impact than school closure in the reduction of fatality rate in all countries, though the school closure was not negligible nor detrimental only in the developing and least developed countries. 

The developing country presented the highest fatality rate, which was 5.3 times higher than the fatality rate in the least developed country. The aged developed country also had a high fatality rate, which was only 6\% lower than the developed country value, so we concluded that the vulnerability due to its older population was somewhat compensated by the lower contact rate between individuals. A remarkable result was that all the intervention policies had a relatively small impact in the fatality rate in the developing country, which we attributed to the high home contact rates between the young and the elderly. This social characteristic is also present in the least developed country, but high fatality rates were not observed due to the small fraction of elderly individuals in this country.

\appendix
\section{Relation between the model parameters and rate coefficients }
\label{appendix}

The current model must predict a number of total deaths and hospitalization which is consistent with the prescribed infection fatality rate and hospitalization rate per age group. For instance, the numerical results must satisfy the conditions
\begin{eqnarray}
T_{L,C}& = & \lim_{t\to\infty} \frac{C_{j}(t)}{R_{j}(t)+C_{j}(t)} \label{cond_one} \\
\phi &= & \lim_{t\to\infty} \frac{\int_{0}^{t} \gamma_H I(t) dt}{\int_{0}^{t} \gamma_{R,I} I(t) dt + \int_{0}^{t} \gamma_H I(t) dt} \label{cond_two} \,\, .
\end{eqnarray} 
From Eq. \ref{cond_two} we may readily derive the expression in Eq. \ref{rel_gammas} by assuming that the hospitalization rate coefficient and the infected recovery rate coefficient are time independent. 

The infection fatality rate (Eq.\ref{cond_one} ) can also be written as 
\begin{equation}
T_{L,C} = \lim_{t\to\infty} \frac{\int_{0}^{t} \mu_{cov} H(t) dt}{\int_{0}^{t} \gamma_H I(t) dt + \int_{0}^{t} \gamma_{R,I} I(t) dt} \,\, .
\end{equation} 
We may use the approximation $\phi \sim H(t)/(I(t)+H(t))$ and express $H(t)$ as a function of $I(t)$. After that simplification and assuming that all parameters are time independent, we may readily derive Eq. \ref{rel_mu}. This expression is not exact, because the relation $\phi \sim H(t)/(I(t)+H(t))$ is an approximation. We inspected numerically the results for our simulation, and we verified that the infection fatality rate predicted by this method is accurate to within a typical error of $10\%$. 

\section{$R0$ calculation}
\label{appendix2}

The basic reproduction number $R0$ is a key parameter in the description of any epidemic, so a deeper analysis of a given mathematical epidemiological model must show how the $R0$ value is computed. The method of evaluation of $R0$ and its formula depends on the model. For instance, an epidemic model based on graphs, such as the one introduced by Oliveira \cite{oliveira2020early}, has a particular formula for $R0$. In the context of epidemic data analysis, there are specific methods to determine $R0$ or $R(t)$ from time series, such as the recent inversion method proposed by Pijpers \cite{pijpers2021non}.  Here, we are interested in showing the mathematical procedure we used to determine the $R0$ value for the present model.  In this case, the next-generation method is the most appropriate tool to compute $R0$ \cite{diekmann1990definition,van2017reproduction}.

The  $\mathbf{F}$ matrix is defined as $F_{i,j}=\partial \mathcal{F}_i/\partial x_j$, where $\mathcal{F}_i$ is the rate of new infections in infectious compartment $i$ and  where $x_{j}$ is any compartment with infected individuals. In our model, there are four higher level infectious compartments, I, A, E and H, and each of them is further divided in 16 age groups, so actually there are 64 compartments. Therefore, $\mathbf{F}$ is a $64 \times 64$ matrix of the following type
\begin{equation}
    \mathbf{F} = \left(
    \begin{array}{@{}c|@{}c|@{}c|@{}c}
     \rule[0pt]{0pt}{\heightof{\bigzero}+1ex} \phantom{_X}\bigF_E  & \phantom{_X}\bigF_I   & \phantom{_X}\bigF_A   & \phantom{_X}\bigzero\phantom{_X} \\[.65ex]
     \hline
    \rule[0pt]{0pt}{\heightof{\bigzero}+1ex} \bigzero  &  \bigzero   & \bigzero & \bigzero \\[.65ex]
    \hline
    \rule[0pt]{0pt}{\heightof{\bigzero}+1ex} \bigzero  &  \bigzero   & \bigzero & \bigzero \\[.65ex]
    \hline
    \rule[0pt]{0pt}{\heightof{\bigzero}+1ex} \bigzero  &  \bigzero   & \bigzero & \bigzero
    \end{array}
    \right)_{64 \times 64} \,\, ,
\end{equation}    
where each matrix block is a $16\times16$ matrix. Note that only the blocks on the fourth column are all zero, since this model neglects new infection by the hospitalized. The non-zero blocks may be explicitly expressed as
\begin{equation}
    \mathbf{F}_E = \left(
    \begin{array}{@{}c@{}c@{}c}
    \frac{\xi_1\beta_{1,1}S_{1}}{N}  & \dots & \frac{\xi_{16}\beta_{1,16}S_{1}}{N} \\
     \vdots & \ddots & \vdots \\
     \frac{\xi_1\beta_{16,1}S_{16}}{N}          & \dots &  \frac{\xi_{16}\beta_{16,16}S_{16}}{N}
    \end{array}
    \right) \,\, ,   
\end{equation}   
\begin{equation}
    \mathbf{F}_I = \left(
    \begin{array}{@{}c@{}c@{}c}
    \frac{\beta_{1,1}S_{1}}{N}  & \dots & \frac{\beta_{1,16}S_{1}}{N} \\
     \vdots & \ddots & \vdots \\
     \frac{\beta_{16,1}S_{16}}{N}          & \dots &  \frac{\beta_{16,16}S_{16}}{N}
    \end{array}
    \right) \,\, ,   
\end{equation}
\begin{equation}
    \mathbf{F}_A = \left(
    \begin{array}{@{}c@{}c@{}c}
    \frac{\alpha_1\beta_{1,1}S_{1}}{N}  & \dots & \frac{\alpha_{16}\beta_{1,16}S_{1}}{N} \\
     \vdots & \ddots & \vdots \\
     \frac{\alpha_1\beta_{16,1}S_{16}}{N}          & \dots &  \frac{\alpha_{16}\beta_{16,16}S_{16}}{N}
    \end{array}
    \right) \,\, .   
\end{equation}   
Let's compute the $\mathbf{V}$ matrix, defined as  $V_{ij}=\partial \mathcal{V}_i/\partial x_j$, where $\mathcal{V}_i$ is  rate of transition of infected to other infection compartments and $x_{j}$ is any compartment with infected individuals. The $\mathbf{V}$ matrix has the following structure,
\begin{equation}
    \mathbf{V} = \left(
    \begin{array}{@{}c|@{}c|@{}c|@{}c}
     \rule[0pt]{0pt}{\heightof{\bigzero}+1ex} \phantom{_X}\bigV_{1,1}  & \phantom{_X}\bigzero\phantom{_X}  & \phantom{_X}\bigzero\phantom{_X}   & \phantom{_X}\bigzero\phantom{_X} \\[.65ex]
     \hline
    \rule[0pt]{0pt}{\heightof{\bigzero}+1ex} \phantom{_X}\bigV_{2,1}  &  \phantom{_X}\bigV_{2,2}   & \bigzero & \bigzero \\[.65ex]
    \hline
    \rule[0pt]{0pt}{\heightof{\bigzero}+1ex} \phantom{_X}\bigV_{3,1}  &  \bigzero   & \phantom{_X}\bigV_{3,3} & \bigzero \\[.65ex]
    \hline
    \rule[0pt]{0pt}{\heightof{\bigzero}+1ex} \bigzero  &  \phantom{_X}\bigV_{4,2}   & \bigzero & \phantom{_X}\bigV_{4,4}
    \end{array}
    \right)_{64 \times 64} \,\, .
\end{equation}    
The non-zero matrix are all diagonal, and they can be explicitly written as
\begin{equation}
    \mathbf{V}_{1,1} = \left(
    \begin{array}{@{}c@{}c@{}c}
    a_1  &  &  \\
      & \ddots &  \\
      &  &  a_{16}
    \end{array}
    \right) \,\, ,   
\end{equation} 
\begin{equation}
    \mathbf{V}_{2,1} = \left(
    \begin{array}{@{}c@{}c@{}c}
    -\rho_1 a_1  &  &  \\
      & \ddots &  \\
      &  &  -\rho_{16} a_{16}
    \end{array}
    \right) \,\, ,   
    \mathbf{V}_{2,2} = \left(
    \begin{array}{@{}c@{}c@{}c}
    \gamma_{1}^{H} + \gamma_{1}^{R,I} &  &  \\
      & \ddots &  \\
      &  & \gamma_{16}^{H} + \gamma_{16}^{R,I}
    \end{array}
    \right) \,\, , \,\,\,\,   
\end{equation} 
\begin{equation}
    \mathbf{V}_{3,1} = \left(
    \begin{array}{@{}c@{}c@{}c}
     - (1 - \rho_1) a_1 &  &  \\
      & \ddots &  \\
      &  & - (1 - \rho_{16}) a_{16}
    \end{array}
    \right) \,\, , \,\,\,\,  
    \mathbf{V}_{3,3} = \left(
    \begin{array}{@{}c@{}c@{}c}
     \gamma_{1}^{R,A} &  &  \\
      & \ddots &  \\
      &  & \gamma_{16}^{R,A}
    \end{array}
    \right) \,\, ,   
\end{equation} 
\begin{equation}
    \mathbf{V}_{4,2} = \left(
    \begin{array}{@{}c@{}c@{}c}
    - \gamma_{1}^{H} &  &  \\
      & \ddots &  \\
      &  & -\gamma_{16}^{H}
    \end{array}
    \right) \,\, , \,\,\,\,  
    \mathbf{V}_{4,4} = \left(
    \begin{array}{@{}c@{}c@{}c}
    \gamma_{1}^{H,R} + \mu_{cov,1}&  &  \\
      & \ddots &  \\
      &  & \gamma_{16}^{H,R} + \mu_{cov,16}
    \end{array}
    \right) \,\, .   
\end{equation} 

Finally, we may compute the $R0$ value as the spectral radius of the matrix $\mathbf{FV}^{-1}$. The matrix $\mathbf{FV}^{-1}$ and its eigenvalues can be expressed analytically, in terms of the model parameters. This could be easily achieved, for instance, using any computational algebra interpreter. However, the matrices are of considerable length, and the final expression is cumbersome, so it is much easier to compute the spectral radius numerically instead of deriving a closed expression ro $R0$ and then compute its value for any given set of parameters.    

\bibliographystyle{spphys}
\bibliography{biblio}

\end{document}